\documentclass[sn-mathphys, Numbered]{sn-jnl}

\usepackage{graphicx}%
\usepackage{multirow}%
\usepackage{amsmath,amssymb,amsfonts}%
\usepackage{amsthm}%
\usepackage{mathrsfs}%
\usepackage{aas_macros}
\usepackage[title]{appendix}%
\usepackage{xcolor}%
\usepackage{textcomp}%
\usepackage{manyfoot}%
\usepackage{booktabs}%
\usepackage{algorithm}%
\usepackage{algorithmicx}%
\usepackage{algpseudocode}%
\usepackage{listings}%

\theoremstyle{thmstyleone}%
\theoremstyle{thmstyletwo}%

\theoremstyle{thmstylethree}%

\usepackage[normalem]{ulem}
\usepackage{color}

\newcommand{\us}{\textmu{-Spec}}

\usepackage{natbib}

\begin{document}

\title[Article Title]{Optimization of an Optical Testbed for Characterization of EXCLAIM \us~Integrated Spectrometers}


\author*[1]{\fnm{Maryam} \sur{Rahmani}}\email{maryam.rahmani@nasa.gov}
\author[1]{\fnm{Emily M.} \sur{Barrentine}}
\author[1]{\fnm{Eric R.} \sur{Switzer}}
\author[1]{\fnm{Alyssa} \sur{Barlis}}
\author[1]{\fnm{Ari D.} \sur{Brown}}
\author[1]{\fnm{Giuseppe} \sur{Cataldo}}
\author[1]{\fnm{Jake A,} \sur{Connors}}
\author[1]{\fnm{Negar} \sur{Ehsan}}
\author[1]{\fnm{Thomas M.} \sur{Essinger-Hileman}}
\author[1,3]{\fnm{Henry} \sur{Grant}}
\author[1]{\fnm{James} \sur{Hays-Wehle}}
\author[1]{\fnm{Wen-Ting} \sur{Hsieh}}
\author[1]{\fnm{Vilem} \sur{Mikula}}
\author[1]{\fnm{S. Harvey} \sur{Moseley}}
\author[1]{\fnm{Omid} \sur{Noroozian}}
\author[1]{\fnm{Manuel A.} \sur{Quijada}}
\author[1]{\fnm{Jessica} \sur{Patel}}
\author[1]{\fnm{Thomas R.} \sur{Stevenson}}
\author[2]{\fnm{Carole} \sur{Tucker}}
\author[1]{\fnm{Kongpop} \sur{U-Yen}}
\author[1,3]{\fnm{Carolyn G.} \sur{Volpert}}
\author[1]{\fnm{Edward J.} \sur{Wollack}}



\affil[1]{\orgname{NASA Goddard Space Flight Center}, \orgaddress{\city{Greenbelt}, \state{MD}, \country{USA}}}

\affil[2]{\orgname{Cardiff University}, \orgaddress{\city{Cardiff},  \state{Wales}, \country{UK}}}


\affil[3]{\orgname{University of Maryland}, \orgaddress{\city{College Park}, \state{MD}, \country{USA}}}









\abstract{We describe a testbed to characterize the optical response of compact superconducting on-chip spectrometers in development for the Experiment for Cryogenic Large-Aperture Intensity Mapping (EXCLAIM) mission. EXCLAIM is a balloon-borne far-infrared experiment to probe the CO and CII emission lines in galaxies from redshift 3.5 to the present. The spectrometer, called \us, comprises a diffraction grating on a silicon chip coupled to kinetic inductance detectors (KIDs) read out via a single microwave feedline. We use a prototype spectrometer for EXCLAIM to demonstrate our ability to characterize the spectrometer's spectral response using a photomixer source. 
We utilize an on-chip reference detector to normalize relative to spectral structure from the off-chip optics and a silicon etalon to calibrate the absolute frequency.}




\maketitle
\vspace{-8mm}
\section{Introduction}\label{Intro}
\vspace{-2mm}
The EXperiment for Cryogenic Large Aperture Intensity Mapping (EXCLAIM) mission is a far-infrared, balloon-borne, cryogenic telescope operating over 420-540\,GHz which will investigate the star formation rate from present day to redshift $z=3.5$ by examining the carbon monoxide (CO) and singly-ionized carbon ([CII]) emission lines. 
\cite{switzer2021experiment, cataldo2020overview, 2020JLTP..199.1027A}. EXCLAIM features six compact and sensitive spectrometers with a resolving power $R= \Delta \lambda/\lambda = 512$. These spectrometers, referred to as \textmu-Spec, are compact, on-chip, Rowland-type, grating spectrometers with a single-crystal silicon (Si) device layer and superconducting niobium (Nb) planar transmission lines which are coupled to aluminum (Al) KIDs\cite{EXCLAIMspectrometerSPIE2020, mirzaei2020mu, EXCLAIMdetectorSPIE2020}. Here we employ a prototype spectrometer, which has a resolving power $R=\lambda / \Delta \lambda = 64$ and operates from 400-600 GHz 
\cite{cataldo2014micro} to demonstrate our ability to characterize the  spectrometer's spectral response using a photomixer source. In this prototype, as in the EXCLAIM design, the sub-millimeter light couples to the testbed optics through an anti-reflection coated silicon lenslet and on-chip broadband (360-650 GHz) slot antenna \cite{EXCLAIMopticsSPIE2020, 2020SPIE11453E..0HE, 2020SPIE11453E..0MM}. This prototype spectrometer includes 48 Al KIDS which have microwave resonance frequencies from $2.8 - 3.0$\,GHz.  
In the subsequent sections, we introduce our test setup, delve into the methods for calibration, and present the characterization results.\\
\vspace{-8mm}
\begin{figure}[h]
\centering
\includegraphics[width=1.0\textwidth]{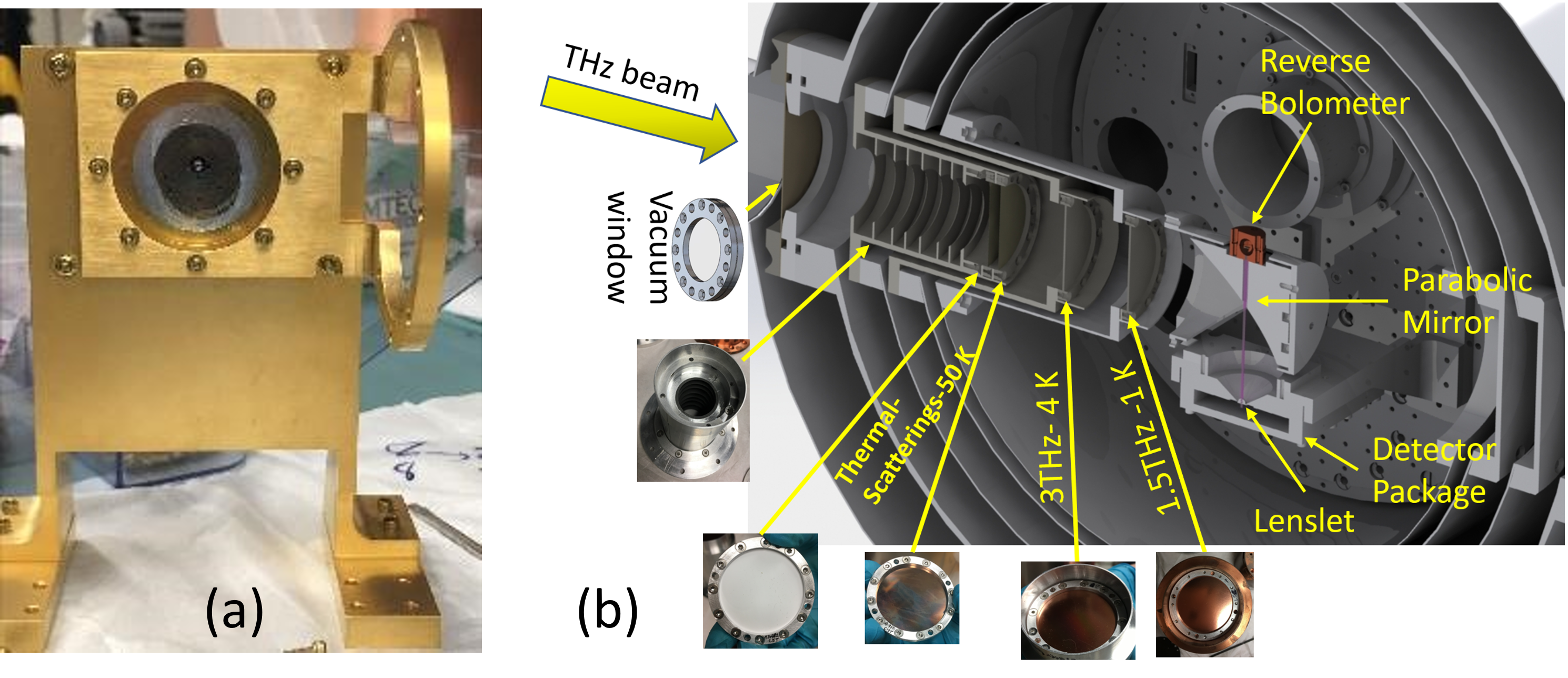}
\vspace{-6mm}
\caption{
(a) The prototype spectrometer mounted prior to installation in the testbed and showing the lenslet aperture side. 
b) The optical window insert in the DR, containing the thermal blocking filters, and a parabolic mirror coupling to the spectrometer.} 
\label{fig:DR_PhMx_BB}
\end{figure}

\vspace{-9mm}

\section{Test Setup}
\label{testbed}
\vspace{-3mm}
The primary testbed for this prototype and the EXCLAIM \us\ is a Blue Fors LD-250 dilution refrigerator (DR) capable of operating at a temperature as low as $7$\,mK. We complete these $R=64$ \us\ measurements 
at $\sim 50$\,mK, which is well below the transition temperature of the superconducting Nb microstrip lines and the Al KIDs. The DR is equipped with two radio-frequency (RF) readout chains, which cover bands at $2-4$\,GHz and $4-8$\,GHz. Each chain also features a cryogenic amplifier and in-line attenuators to regulate the power received by the KIDs and dissipated at each temperature stage of the DR. Optical characterization is provided by either a blackbody source located inside the DR or via coupling optics to an external room-temperature photomixer source. The photomixer enables characterization of the spectral response, and the blackbody provides an absolute power reference to enable measurement of the sensitivity and efficiency. In this paper, we describe the photomixer configuration. 

\vspace{-2mm}
\subsection{The Photomixer and Optical Coupling}
\vspace{-2mm}

We use a continuous-wave Terahertz $780$\,nm Toptica photomixer. The photomixer is constructed from GaAs and operates using the beat frequency of two near-infrared lasers. This provides a narrow-band optical output tunable over $0-1.8$\,THz and well-suited for high-resolution spectroscopy. The source can be modulated by an AC bias up to $48.8$\,kHz to enable synchronous, lock-in detection. The output of the photomixer passes through the DR's radiation shells via an optical path containing thermal blocking filters (see Fig.\,\ref{fig:DR_PhMx_BB}). 
Characterization of the spectrometer's spectral response is implemented by sweeping the frequency of the photomixer signal and measuring the KID response of each spectrometer channel.

\vspace{-2mm}
\subsection{Readout}
\vspace{-2mm}
 The main components of the room temperature KID microwave readout are a Reconfigurable Open Architecture Computing Hardware (ROACH2) system, a digital signal processing (DSP) and data analysis/acquisition platform and an IF-box. In our testbed, the ROACH2 is used to generate a comb of frequency tones to simultaneously read out the 48 resonators on the prototype \us, and ultimately 355 resonators in EXCLAIM's \us. 
 The IF box is responsible for upconverting and downconverting the 48 tones from the ROACH2 digital-to-analog converter (DAC), which has 500 MHz bandwidth, to the resonance frequencies of the KIDs at $2.0 - 2.5$\,GHz using mixers and a local oscillator (LO) signal generated by a synthesizer.
This signal is then sent into the DR RF chain and to the on-chip KID feedline. When an optical signal is absorbed by a KID, the resonance frequency slightly changes and the amplitude and phase of the transmitted RF signal also change. This RF signal is returned from the DR to the IF-box and demodulated with the same LO signal to generate an I/Q signal which is then sent back to the analog-to-digital (ADC) block of the ROACH2. We also optimize the IF-box gain and attenuation configuration. By 
 adjusting the gain and attenuation levels, we can maximize the signal-to-noise (SNR).
 
\vspace{-2mm}
\section{Optical Characterization Procedure}\label{optimization}
\vspace{-2mm}
To optimize our measurements we take several measures that are described below.
\vspace{-2mm}
\subsection{Reference Detector}
\vspace{-2mm}
The prototype spectrometer includes a microstrip Wilkinson power divider located after the slot antenna, but prior to the input to the spectrometer. This divider directs half of the signal into the spectrometer and half to a ``reference" KID located on the same microwave feedline as the spectrometer KIDs. A similar reference detector (coupled via a -20 dB coupler, so as to have negligible impact on the spectrometer's optical efficiency) is included in the EXCLAIM spectrometer design for this purpose. 
Alignment of the collimated photomixer beam with the vacuum window of the DR and the \us\ beam is implemented by maximizing the photomixer signal response of the broadband reference detector in real-time by translating a moving stage which houses the photomixer.
During an optical frequency scan, the reference KID also provides a way to simultaneously monitor the optical response at the input to the spectrometer. Structure that is observed in the response of both the spectrometer and reference KIDs, as the photomixer source is scanned in optical frequency, and which may be due to reflections in the off-chip optics, can be separated from the intrinsic spectrometer grating response by dividing each KID response by the reference KID response. 

\vspace{-2mm}
\subsection{Absolute Frequency Calibration}
\begin{figure}[ht!]
 \centering
\includegraphics[width=1.0\textwidth]{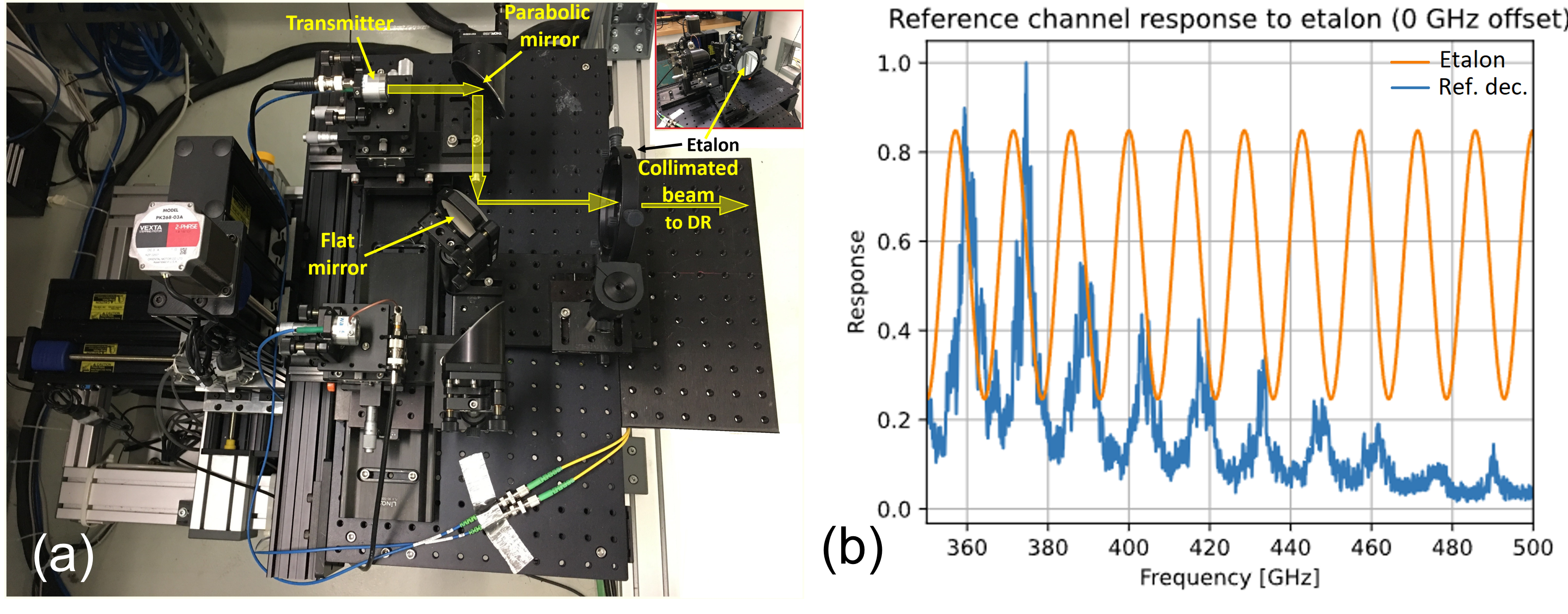}
\vspace{-6mm}
\caption{ a) The etalon position on the photomixer optical stage (the red box at top right shows another perspective). b) A comparison of the etalon's fringes in the reference channel response compared to their nominal frequency locations from the photomixer scan parameters.}
\label{fig:etalon}
\end{figure}

 The frequency of the laser beams generated by the photomixer is susceptible to temperature drifts, and the absolute frequency stability is only guaranteed to within $2$~GHz, with a relative frequency stability of $\le10$ MHz in ideal conditions \cite{Toptica}. EXCLAIM's science goals require an absolute frequency calibration within $\pm0.1$~GHz. Consequently, optical measurements of the spectral response of the \us\ require additional calibration to achieve the necessary absolute frequency knowledge, and the system is calibrated using an etalon. The etalon used is a Tydex high resistivity float-zone Si Fabry-Perot etalon (model HRFZ-Si-D25.4-T3) which has a diameter of $25.4$~mm, thickness of $3$~mm, a free spectral range (FSR) of $14.5$~GHz, a full width at half maximum (FWHM) of $5.5$~GHz, and a finesse of $2.6$. An etalon consists of two parallel surfaces of a silicon plate that form a resonant cavity. A simplified equation for calculating the resonance frequency of an etalon is $f_r = c/(2nL)$. In this equation, $c$ is the speed of light in vacuum, $n$ the refractive index of the etalon's material, $L$ is the thickness of the etalon cavity. However, since the etalon thickness is not determined accurately enough by the manufacturer (the thickness is measured to 0.001\,mm precision) to provide the absolute frequency accuracy needed instead we measure the etalon's spectral response directly using a Bruker 125HR model Fourier Transform Spectrometer with a spectral resolution of $0.05$ icm (from 15-670 icm). These measurements are shown in Fig.~\ref{fig:etalon}.
 
 The etalon is inserted at a normal angle to the photomixer beam prior to the beam entering the cryostat (see Fig.~\ref{fig:etalon}). The etalon must be aligned within $7^{\circ}$ to achieve $\pm 0.1$~GHz, which is done using reflection from a laser in a metrology jig along the optical axis (see Fig.~\ref{fig:etalon}) When the photomixer beam is incident on the etalon, it forms a series of equally spaced resonant modes. By observing this frequency spacing in the broadband reference signal, the absolute frequency of the photomixer beam incident on the spectrometers is determined. An optical frequency scan is completed by the photomixer with and without the etalon in the photomixer optical path. The raw time stream data is first converted to the optical frequency determined by the photomixer controller and nominal frequency scan parameters. Then the spectral response of the reference resonator is compared to the etalon spectral response and the optical frequency values are shifted by the offset observed. 
 
\vspace{-4mm}

\vspace{+2mm}
\subsection{Signal Processing}
\vspace{0mm}
We use custom data acquisition (DAQ) Python code to process the data.
Modulation of the photomixer source enables the detection system to be tuned to the modulation frequency thereby isolating the signal of interest from slow drifts (so-called ``$1/f$'' noise). This acts as a lock-in-amplifier, which is capable of extracting the signal of interest from a noisy background by precisely synchronizing with the modulation frequency. During the signal processing, the mean value of the signals are subtracted to remove the DC component in the frequency domain in order not to mask the signal. We use a Fourier transform to cross-correlate each spectrometer channel and the reference channel, and then find the signal power at the photomixer modulation frequency. We then normalize by the power measured in the reference channel. The KID resonator ring down time, determined by the resonator quality factor and resonance frequency ($\tau = Q/\pi f_0$), and the quasiparticle recombination rate may limit the modulation frequency. The ROACH2 digital electronics sampling rate of 1024\,Hz sets the upper limit to this modulation frequency. The optimization of these factors led to the selection of a modulation frequency of $200$~Hz for our tests.

Another parameter affecting the SNR is the integration time. By extending the averaging time, the impact of noise can be minimized. Over a longer integration time, although both signal and noise accumulate, due to the random and uncorrelated nature of the noise, it averages out over time and a higher $SNR$ is achieved. Furthermore, increasing the integration time yields more data points and improves the statistical robustness of the measurement.

 \vspace{-2mm}
\section{Measurement Results}\label{results}
\begin{figure}[ht!]
 \centering
\includegraphics[width=1.0\textwidth]{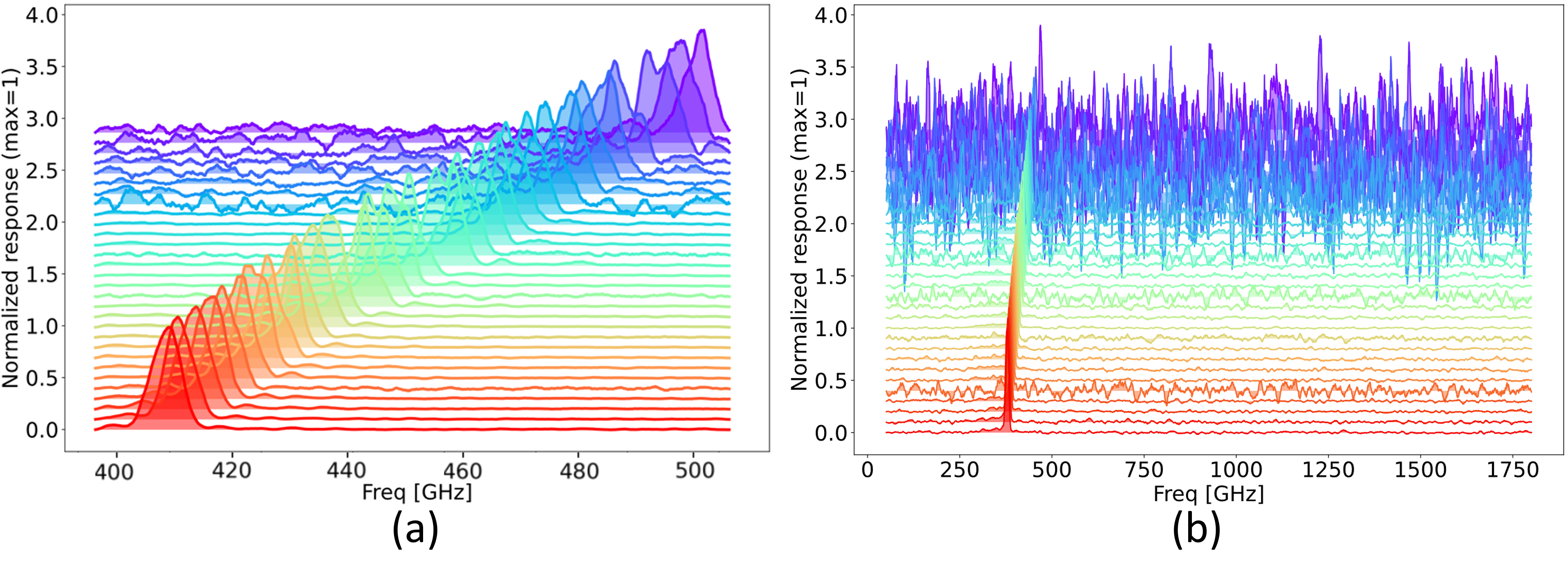}
\vspace{-8mm}
\caption{The prototype spectrometer response (a) across the spectrometer band and b) the out-of-band spectral response. } \label{fig:R64performance}
\end{figure}

\vspace{-2mm}
Work is on-going to characterize the absolute frequency response, resolution and out-of-band response of the prototype $R=64$ spectrometer. Notably, the reference channel does exhibit a $4$\,GHz frequency difference (error) with respect to the etalon fringes (see Fig.~\ref{fig:etalon}), which is twice the specified accuracy of $2$\,GHz absolute frequency, and may be due to non-ideal control of the laboratory temperature. The iteration of the measurement with the etalon under the same settings and lab environmental conditions revealed a $1$~GHz relative frequency difference between measurements and an additional $1$~GHz frequency drift for frequencies toward the high end of the band $\sim 500$~GHz. This necessitates additional refinement of the frequency data across the scans to calibrate within EXCLAIM's $0.1$ GHz frequency accuracy requirement. Given the specifications of the etalon and photomixer discussed earlier, we should be able to achieve the required absolute frequency accuracy. 
Fig.~\ref{fig:R64performance}-a depicts the response of thirty of the spectrometer channels from $412 - 505$~GHz. In this measurement, the photomixer source was modulated at $f_m = 200$~Hz, 
and the optical frequency step size was $\Delta \nu = 0.02$~GHz. Fig.~\ref{fig:R64performance}-b shows the out of band spectral response of the R64 \us.

\vspace{-3mm}

\vspace{-8mm}
\section{Conclusion}
\vspace{-2mm}
In this work, we described the primary methods for spectral characterization of \us\ on-chip spectrometers. We introduced the main components of this setup, which are the photomixer source, etalon, reference detector and readout system. We discussed the use of a Si etalon to calibrate the photomixer and the spectrometer's absolute frequency response as well as the optimization of the scan parameters for increased signal to noise. 
The methods described here will be used in the future optical characterization of the higher-resolution EXCLAIM spectrometers.

\vspace{-2mm}
\section{Acknowledgment}
\vspace{-2mm}
The authors acknowledge the EXCLAIM funding (No. APRA 17-APRA17-0077) grant, prior funding through the ROSES APRA program for the prototype spectrometer development, and an Oak Ridge Associated Universities (ORAU) fellowship.


\bibliography{main}


\end{document}